\documentclass[]{article}
\usepackage[left=3cm, right=2cm, top=2cm]{geometry}
\usepackage{graphicx}
\usepackage{graphicx,epstopdf}
\usepackage{float}
\usepackage{amssymb,amsmath}
\usepackage{exscale}
\usepackage{makeidx,shortvrb,latexsym}
\usepackage{epstopdf}
\usepackage{tabularx, booktabs, multirow}
\usepackage{array}
\usepackage[]{hyperref}
\usepackage{color}
\usepackage{authblk}
\usepackage{natbib}

\begin{document}

\title{An algorithm for stress and mixed control in Galerkin based FFT homogenization}

\author[1,2]{S. Lucarini}

\author[1,2,*]{J. Segurado}

\affil[1]{IMDEA Materials Institute ,C/ Eric Kandel 2, 28906, Getafe, Madrid, Spain.}

\affil[2]{Technical University of Madrid/Universidad Polit\'ecnica de Madrid, Department of Materials Science, E. T. S. de Ingenieros de Caminos \\C/ Profesor Aranguren s/n 28040 - Madrid, Spain.}

\affil[*]{Corresponding author at: IMDEA Materials Institute, Spain. javier.segurado@imdea.org (J. Segurado)}


\maketitle

\begin{abstract}
A new algorithm is proposed to impose a macroscopic stress or mixed stress/deformation gradient history in the context of non-linear Galerkin based FFT homogenization. The method proposed is based in the definition of a modified projection operator in which the null frequencies enforce the type of control (stress or strain) for each component of either the macroscopic first Piola stress or the deformation gradient. The resulting problem is solved exactly as the original variational method and it does not require additional iterations compared to the strain control version, neither in the linear iterative solver, nor in the Newton scheme. The efficiency of the method proposed is demonstrated with a series of numerical examples, including a polycrystal and a particle reinforced hyperelastic material.
\end{abstract}
\footnote{Submitted to International Journal of Numerical Methods is Engineering}

\section{Introduction}

Computational homogenization aims to simulate the mechanical response of a material under a given macroscopic stress, strain or mixed loading path by solving a boundary problem in a periodic representative volume element (RVE) of its microstructure. The finite element method (FE) has been traditionally used to solve the boundary value problem allowing to impose any macroscopic load history combining the control of different components of the strain and stress average fields. Fast Fourier transform (FFT) based homogenization has become a very popular alternative to FE. FFT methods solve the boundary problem replacing the differential operators in the governing partial differential equation by their equivalent counterparts in the Fourier space. The computational performance of this method compared to FE makes it ideal to simulate complex geometries comprising millions of degrees of freedom. However, imposing a macroscopic  stress or mixed loading path is not a direct task in FFT. Nevertheless, obtaining the material response in an efficient manner for any generic loading path is fundamental to solve cases of interest such as uniaxial loading, controlled triaxility or cyclic stress control tests.

FFT homogenization was first proposed by H. Moulinec and P. Suquet \citep{Suquet1994,Suquet1998} based on solving the Lippmann-Schwinger equation with a reference material using a fixed-point algorithm. Since then, several approaches based on similar principles have been proposed  \citep{Milton1999,Michel2001,Monchiet2012,Kabel2014} . These approaches require the use of a reference medium, which properties in some cases may strongly determine the convergence rate. In these \emph{classical schemes}, load is introduced imposing the full macroscopic strain tensor and when stress or mixed control is required (i.e. to impose uniaxial tension) iterative approaches are the usual solution \cite{Michel1999,Eisenlohr2013} . Under this procedure the algorithm still imposes the full macroscopic strain tensor and corrects it in an iterative manner in order to obtain at the end of the iteration process the macroscopic stress tensor components imposed as input. As a result of the corrections to the applied strain during iterations, the number of iterations per step significantly increases respect the strain control. An alternative approach for classic schemes under mixed control was recently proposed by Kabel et al. \citep{Kabel2015}. The method consists in modifying the original basic scheme \citep{Suquet1994} through the introduction of projection tensors to account for the load control and solving the problem using an iterative Krylov solver. This method for stress control is more efficient than the iterative approaches although still requires the definition of a reference medium which properties might influence the convergence rate.

A different approach for FFT homogenization is the Galerkin based approach in which the problem is derived using a weak formulation, very similar to finite elements, and projection operators that ensure the compatibility of the strain fields. The resulting method \citep{Vondrejc2014,Geers2016,Geers2017} does not require a reference medium and make direct use of the corresponding tangents of the material behavior to improve the convergence. Moreover, this framework can be easily adapted to use any complex non-linear constitutive equation as the crystal plasticity model \citep{Lucarini2018}. However, the variational approach was still formulated imposing the history of the macroscopic deformation gradient.

In this work, we extend the variational algorithm for finite strains\citep{Geers2017} to use stress and mixed control of the macroscopic load history using a direct method based on a modified project operator and preserving the computational performance of the original method. This paper summarizes in Section 2 the original FFT-based variational approach and then presents the modified projection operator and the resulting algorithm for mixed control. In Section 3 examples using different material models and loading paths will be run and, finally, the performance of the method is analyzed.

\section{Algorithm for stress and mixed control in Galerkin based FFT}
\subsection{Galerkin based FFT for strain control}
The aim of the homogenization method is to find, for a given macroscopic load history, the value of the deformation gradient and stress on a periodic RVE that can for example consist of a heterogeneous material composed by the spatial arrangement of two or more phases with different properties or constitutive laws. The RVE is discretized in $n_x \ n_y \ n_z$ voxels. The linear momentum balance is imposed in its weak form in the reference configuration. The test functions are virtual deformation gradients premultiplied by a projector operator to enforce their compatibility. After some algebraic work \citep{Geers2017}, the discrete form of the linear momentum balance can be written as
\begin{equation}
\mathcal{G}\left(\mathbf{P}\left(\mathbf{F},\boldsymbol{\alpha}\right)\right)=\mathbf{0} \label{eql_discrete1}
\end{equation}
where $\mathbf{P}$ is the first Piola Kirchhoff stress, $\mathbf{F}$ is the deformation gradient field, $\mathbf{\alpha}$ are a set of internal variables defining the history and $\mathcal{G}$ is a linear map that acts on a voxel tensor field $\mathbf{P}$ defined in $\mathbb{R}^{9 \ n_x \ n_y \ n_z}$ to create a new voxel tensor field defined in the same space. The linear map corresponds to a convolution in the real space that is computed as
\begin{equation}\mathcal{F}^{-1}\left\{ \widehat{\mathbb{G}}:\mathcal{F} \left\{\mathbf{P}\left(\mathbf{F},\boldsymbol{\alpha}\right) \right\} \right\}=\mathbf{0}
\label{eql_discrete2}
\end{equation}
where $\mathcal{F}$ and $\mathcal{F}^{-1}$ are the discrete Fourier transform and its inverse and $\hat{\mathbb{G}}$ is the projection operator in the Fourier space (fourth order tensor field). Equation (\ref{eql_discrete2}) is an algebraic system of $9\ n_x \ n_y \ n_z$ equations in which the unknown is the value of the deformation gradient at each voxel $\mathbf{F}$ which can be decomposed into the sum of its average value $\left<\mathbf{F} \right>_{\Omega}$ and the fluctuation, $\widetilde{\mathbf{F}}=\mathbf{F}-\left<\mathbf{F} \right>_{\Omega}$. In the case of strain control, the value of $\overline{\mathbf{F}}=\left<\mathbf{F} \right>_{\Omega}$ is given. This boundary condition enters explicitly in the system and the resulting problem consists in finding the value of the fluctuations of the deformation gradient that fulfills
\begin{equation}
\text{Find }\widetilde{\mathbf{F}} \mid \ \mathcal{G}\left( \mathbf{P}\left( \overline{\mathbf{F}}+\widetilde{\mathbf{F}},\boldsymbol{\alpha} \right)  \right)=\mathbf{0}    \ \text{for a given} \ \overline{\mathbf{F}}
\label{eql_discrete3}
\end{equation}

In the case of non-linear material behavior the discrete expression of the weak form of equilibrium given by equation (\ref{eql_discrete2}) defines a non-linear algebraic equation.  In order to solve the non-linear problem, the macroscopic deformation gradient is usually split in $n$ increments and a non-linear equation is solved for every time increment $k\in [1,n]$. Following this approach, the non-linear system of equations defined by equation (\ref{eql_discrete2}) is solved iteratively linearizing in each increment the stress around the last iteration of the deformation gradient, $\mathbf{F}^i$,
\begin{equation}
\mathbf{F}^{i+1}=\mathbf{F}^{i}+{\delta}\mathbf{F} \text{ and } \mathbf{P}(\mathbf{F}^{i}+{\delta}\mathbf{F})\approx\mathbf{P}(\mathbf{F}^{i})+\mathbb{K}^{i}:\delta \mathbf{F} \text{, with } \mathbb{K}^{i}=\frac{\partial \mathbf{P}}{\partial \mathbf{F}^{i}}
\label{eq:lineariz}
\end{equation}
Combining the equilibrium equation (\ref{eql_discrete2}) with the linearization of the stress, the next equation is derived
\begin{equation}\label{linequil}
\mathcal{G}\left( \mathbb{K}^i:\delta \mathbf{F}\right)=-\mathcal{G}\left(\mathbf{P}\left(\mathbf{F}^{i}\right)\right)
\end{equation}
The expression in equation (\ref{linequil}) is a linear system of equations in which the unknown is the correction at the iteration $i$ of the deformation gradient $\delta \mathbf{F}$ obtained in a previous iteration, $\mathbf{F}^{i}=\overline{\mathbf{F}}_{k+1}+\delta \mathbf{F}^1+\delta \mathbf{F}^2\cdots+\delta \mathbf{F}^{i-1}$. All the corrections $\delta \mathbf{F}$ are fluctuation terms (average free) and the applied deformation gradient enters through $\overline{\mathbf{F}}_{k+1}$. The linear equations are solved using the conjugate gradient method. If the constitutive equations are rate-dependent, the stress depends on both the deformation gradient, $\mathbf{F}$, and the velocity gradient, $\mathbf{L}$. In this case the macroscopic strain history is divided in $n$ time increments, corresponding the end of the increment $k$ to a time $t_k$. In each point $\mathbf{F}_{k}=\mathbf{F}(t=t_k)$ and the velocity gradient $\mathbf{L}_k$ is given by
$$\mathbf{L}_k=\frac{1}{t_{k}-t_{k-1}}[\mathbf{F}_k-\mathbf{F}_{k-1}]\cdot\mathbf{F}_k^{-1}$$

\subsubsection*{Projection operator}
The projection operator $\widehat{\mathbb{G}}\left(\mathbf{x}\right)$ is a fourth order tensor (with a closed form expression in the Fourier space) that enforces the compatibility of any tensor field via convolution in the real space. In this framework, the projection operator is used to force the compatibility of the deformation gradient tensor field.

The Helmholtz decomposition of an arbitrary tensor field $\mathbf{A}\left(\mathbf{x}\right)$ allows to express the field as the sum of three contributions
\begin{equation}
\mathbf{A}\left(\mathbf{x}\right)=\mathbf{A}_\parallel \left(\mathbf{x}\right)+\mathbf{A}_\perp\left(\mathbf{x}\right)+\overline{\mathbf{A}}
\label{helm}
\end{equation}
where $\mathbf{A}_\parallel\left(\mathbf{x}\right)$ is a curl-free component (compatible), $\mathbf{A}_\perp\left(\mathbf{x}\right)$ a divergence-free component (incompatible) and  $\overline{\mathbf{A}}$ is the average. The Helmholtz decomposition in the Fourier space \citep{Stewart2012} is written as
\begin{equation}
\widehat{\mathbf{A}}\left(\boldsymbol{\xi}\right)=\widehat{\mathbf{A}_\parallel}\left(\boldsymbol{\xi}\right)+\widehat{\mathbf{A}_\perp}\left(\boldsymbol{\xi}\right)+\widehat{\overline{\mathbf{A}}},
\label{helmfour}
\end{equation}
being the curl-free component 
\begin{equation}
\widehat{\mathbf{A}_\parallel}\left(\boldsymbol{\xi}\right)=\left(\widehat{\mathbf{A}}\left(\boldsymbol{\xi}\right)\cdot\frac{\boldsymbol{\xi}}{\lvert\boldsymbol{\xi}\rvert}\right)\otimes\frac{\boldsymbol{\xi}}{\lvert\boldsymbol{\xi}\rvert}
\label{comp}
\end{equation}
where $\boldsymbol{\xi}$ represents the (spatial) frequency vector, with dimension of inverse of length \cite{Lucarini2018}. A fourth order tensor, the projection operator $\widehat{\mathbb{G}}$, can be defined from equation (\ref{comp}) which, when contracted with $\widehat{\mathbf{A}}$, returns its compatible part
\begin{equation}
\widehat{\mathbb{G}}:\widehat{\mathbf{A}}=\widehat{G}_{ijkl}\widehat{A}_{kl}=\widehat{A}_{\parallel_{ij}}=\delta_{ik}\frac{\xi_j\xi_l}{\boldsymbol{\xi}\cdot\boldsymbol{\xi}} \widehat{A}_{kl}=\frac{\xi_j\xi_l}{\boldsymbol{\xi}\cdot\boldsymbol{\xi}} \widehat{A}_{il}
\end{equation}
In  this expression it can be observed that the projection operator in the Fourier space accomplish major symmetry $\widehat{G}_{ijkl}=\widehat{G}_{klij}$.

In order to return a zero mean field value, the projection should be a zero fourth order tensor in the null frequency
 $\boldsymbol{\xi}=\left(0,0,0\right)$. Additionally, when discretized in an even number of frequencies, the value of the operator at the highest frequency (Nyquist frequency) is usually approximated to a null fourth order tensor to recover the frequency symmetry. Considering this, the final expression for the projector operator is
\begin{equation}
\widehat{\mathbb{G}}\left(\boldsymbol{\xi}\right)=\widehat{G}_{ijkl} = \left\{ \begin{array}{lr} 0_{ijkl} & \text{for null and Nyquist frequencies} \\ \delta_{ik} \frac{\xi_j \xi_l}{\boldsymbol{\xi}\cdot\boldsymbol{\xi}} & \text{for the rest of frequencies} \end{array} \right.
\label{eq:G}
\end{equation}
which corresponds to the Green Function of the \emph{basic scheeme} \cite{Suquet1994} for an isotropic reference medium with Lam\'e constants $\lambda=0,\mu=0.5$.

\subsection{Macroscopic stress control}
The Galerkin based FFT method \citep{Vondrejc2014,Geers2016,Geers2017} was originally developed to provide a solution under a given macroscopic deformation gradient. In this section, a modification of the projection operator is proposed in order to simulate the behavior for a macroscopic stress history. This case would correspond to a creep test or a unixial test under stress control.

The first Piola-Kirchhoff stress can be split into its average and fluctuating parts
\begin{equation}
\mathbf{P}\left(\mathbf{F},\boldsymbol{\alpha}\right)=\left<\mathbf{P}(\mathbf{x},\boldsymbol{\alpha})\right>_\Omega +\widetilde{\mathbf{P}}(\mathbf{F},\boldsymbol{\alpha}) 
\label{eqP1}
\end{equation}
and introducing this decomposition into equilibrium equation (\ref{eql_discrete1}), it is obtained 
\begin{equation}
\mathcal{G}\left(\left<\mathbf{P}\right>_\Omega+\widetilde{\mathbf{P}}\left(\mathbf{F},\boldsymbol{\alpha}\right)\right)=\mathbf{0}
\end{equation}
The average term of $\mathbf{P}$ vanishes because the projection operator $\mathbb{\hat{G}}$ eliminates the mean part of the field to which is applied and the actual equilibrium equation that is solved corresponds to
\begin{equation}
\mathcal{G}\left(\widetilde{\mathbf{P}}\left(\mathbf{F},\boldsymbol{\alpha}\right)\right)=\mathbf{0} \ \mathrm{for} \ \left<\mathbf{F}\right>_\Omega=\overline{\mathbf{F}} 
\label{eqP2}
\end{equation}
where the input data is the averaged value of the deformation gradient and the average stress is unknown and obtained as a result of the simulation.

Let's consider and alternative projection operator $\mathcal{G}^\ast$ in which the condition of zero mean projection is eliminated. If this operator is applied to the fluctuation of the stress (zero mean value), the redefinition of the operator does not affect the result and the equilibrium equation is identical to the one expressed using the standard operator (eq. \ref{eqP2}) coincides 
\begin{equation}
\mathcal{G}^\ast\left(\widetilde{\mathbf{P}}\left(\mathbf{F},\boldsymbol{\alpha}\right)\right)=\mathcal{G}\left(\widetilde{\mathbf{P}}\left(\mathbf{F},\boldsymbol{\alpha}\right)\right)=\mathbf{0}.
\label{eqP3}
\end{equation}
Under stress control the input data is the target macroscopic stress, $\overline{\mathbf{P}}$, and therefore it should be equaled to the average stress $\left<\mathbf{P}\right>_\Omega$. Combining the stress decomposition (equation \ref{eqP1}) with equation (\ref{eqP3}), the equilibrium can be expressed as
\begin{equation}
\mathcal{G}^\ast\left(\mathbf{P}\left(\mathbf{F},\boldsymbol{\alpha}\right)-\left<\mathbf{P}\right>_\Omega\right)=\mathcal{G}^\ast\left(\mathbf{P}\left(\mathbf{F},\boldsymbol{\alpha}\right)-\overline{\mathbf{P}}\right)=\mathbf{0} 
\end{equation}
The resulting equilibrium equation consists in finding the value of the full deformation gradient, $\mathbf{F}$, including both fluctuations and the field mean value and corresponds to
\begin{equation}
\text{Find }\mathbf{F} \mid \mathcal{G}^\ast\left(\mathbf{P}\left(\mathbf{F},\boldsymbol{\alpha}\right)\right)=\mathcal{G}^\ast(\overline{\mathbf{P}}) \text{ for a given $\overline{\mathbf{P}}$}
\label{eqP5}
\end{equation}
To determine the value of $\mathcal{G}^\ast$ that preserves equilibrium enforcing that $\left<\mathbf{P}\right>_\Omega=\overline{\mathbf{P}}$, equation (\ref{eqP5}) is expressed as a convolution in the Fourier space
\begin{equation}
\mathcal{F}^{-1} \left( \widehat{\mathbb{G}^\ast}(\boldsymbol{\xi})  :  \widehat{\mathbf{P}}(\boldsymbol{\xi}) \right) = \mathcal{F}^{-1} \left( \widehat{\mathbb{G}^\ast}(\boldsymbol{\xi})   :  \widehat{\mathbf{\overline{P}}} \right).
\label{eqP7}
\end{equation}
Equation (\ref{eqP3}) implies that both operators should act identically on zero average tensor so $\widehat{\mathbb{G}^\ast}(\boldsymbol{\xi})=\widehat{\mathbb{G}}(\boldsymbol{\xi}) \ \forall \xi\neq 0$, reducing eq. (\ref{eqP7}) to
\begin{equation}
\widehat{\mathbb{G}^\ast}(\boldsymbol{\xi=0}): \widehat{\mathbf{P}}(\boldsymbol{\xi=0})=\widehat{\mathbb{G}^\ast}(\boldsymbol{\xi=0}): \mathbf{\overline{P}}
\label{eq:G0freq}
\end{equation}
Finally, the $\mathbb{G}^*$ operator projects an arbitrary tensor field into the sum of its compatible and mean parts. Therefore, the null frequency should project the average part of a field into itself, and this is achieved by defining its value as the fourth order identity tensor. Using the property of the Fourier transform  $\widehat{\mathbf{P}}(\boldsymbol{\xi=0})=\left<\mathbf{P}\right>_\Omega$ in equation (\ref{eq:G0freq}) , it is enforced $\left<\mathbf{P}\right>_\Omega$ to be the target macroscopic stress. The new operator is defined as
\begin{equation}
\widehat{G}_{ijkl}^{\ast} = \left\{ \begin{array}{lr} \delta_{ik}\delta_{jl} & \text{for null frequency} \\ 0_{ijkl} & \text{for Nyquist frequencies} \\ \delta_{ik} \frac{\xi_j \xi_l}{\boldsymbol{\xi}\cdot\boldsymbol{\xi}}  & \text{for the rest of frequencies} \end{array} \right.
\label{eq:Gnew}
\end{equation}

In the case of non-linear behavior, the system is also solved iteratively. Each iteration for the correction of the deformation gradient $\delta \mathbf{F}$ is obtained here solving this linear system 
\begin{equation}\label{linequil2ast}
\mathcal{G}^\ast\left(\mathbb{K}:\delta \mathbf{F}\right)=-\mathcal{G}^\ast\left(\mathbf{P}\left(\mathbf{F}^{i}\right)-\overline{\mathbf{P}}\right)
\end{equation}
where, contrary to the strain control approach, the correction for the deformation gradient field $\delta \mathbf{F}$ includes a mean value. It can also be observed that when equilibrium is reached ($\delta \mathbf{F}\rightarrow 0$) the average stress reaches the objective. It will be shown in the numerical examples that the algorithm summarized in equation (\ref{linequil2ast}) has the same convergence rate than the original scheme.

\subsection{Mixed control of macrosopic stress and strain}
Load histories combining values of both macroscopic stress and deformation gradient are very common.  Such conditions can be found for example in uniaxial or biaxial tests under strain where strain is imposed in some directions, but leaving the others free to deform. The input data now is $\overline{P}_{IJ}=\left<P\right>_{IJ}$ for components $IJ$ in which stress is imposed and $\overline{F}_{ij}=\left<F\right>_{ij}$ for the rest of components. The equilibrium equation can be expressed as 
\begin{equation}\label{eq:mixedeqil}
\text{Find } \mathbf{F}' \mid \left\{ 
\begin{array}{c}
\mathcal{G}^\ast\left(\mathbf{P}\left(\overline{\mathbf{F}}_{ij}+\mathbf{F}',\boldsymbol{\alpha}\right)\right)_{ij}=0_{ij} \ \text{for a given $\overline{F}_{ij}$ in directions $ij$} \\
\mathcal{G}^\ast\left(\mathbf{P}\left(\overline{\mathbf{F}}_{ij}+\mathbf{F}',\boldsymbol{\alpha}\right)\right)_{IJ}=\mathcal{G}^\ast(\overline{\mathbf{P}})_{IJ} \ \text{for a given $\overline{P}_{IJ}$ in directions $IJ$}
\end{array}
\right.
\end{equation}
where $\mathbf{F}'$ is the solution of the equilibrium equation under mixed control (eq. \ref{eq:mixedeqil}) and corresponds to a fluctuation term for the $ij$ components ($F'_{ij}=\widetilde{F}_{ij}$) and the full deformation gradient term for the $IJ$ components ($F'_{IJ}=F_{IJ}$). These conditions are imposed again by modifying the value of the projection operator for the null frequency. The modified projection operator is obtained following the same reasoning as for stress control and corresponds to
\begin{equation}
\widehat{G}_{ijkl}^{\ast} = \left\{ 
\begin{array}{lr}
\delta_{ik}\delta_{jl} &   \text{if }\mathbf{\xi=0} \text{ for stress controled $IJ$ terms}  \\ 
0_{ijkl} &  \text{if }\mathbf{\xi=0} \text{ for strain controlled $ij$ terms} \\
0_{ijkl} & \text{for Nyquist frequencies} \\
\delta_{ik} \frac{\xi_j \xi_l }{\boldsymbol{\xi}\cdot\boldsymbol{\xi}}  & \text{for } \mathbf{\xi\neq0}
\end{array} 
\right.
\label{eq:Gnewnew}
\end{equation}
This operator should be used as the operator $\mathcal{G}^\ast$ in equation (\ref{eq:mixedeqil}). The resulting non linear system for obtaining the equilibrium is solved by Newton as in the previous cases, and the result is the field $\mathbf{F}'$
\section{Results}\label{examples}
\subsection{Particle reinforced hyperelastic matrix}
\begin{figure}[h]
\includegraphics[width=.32\textwidth]{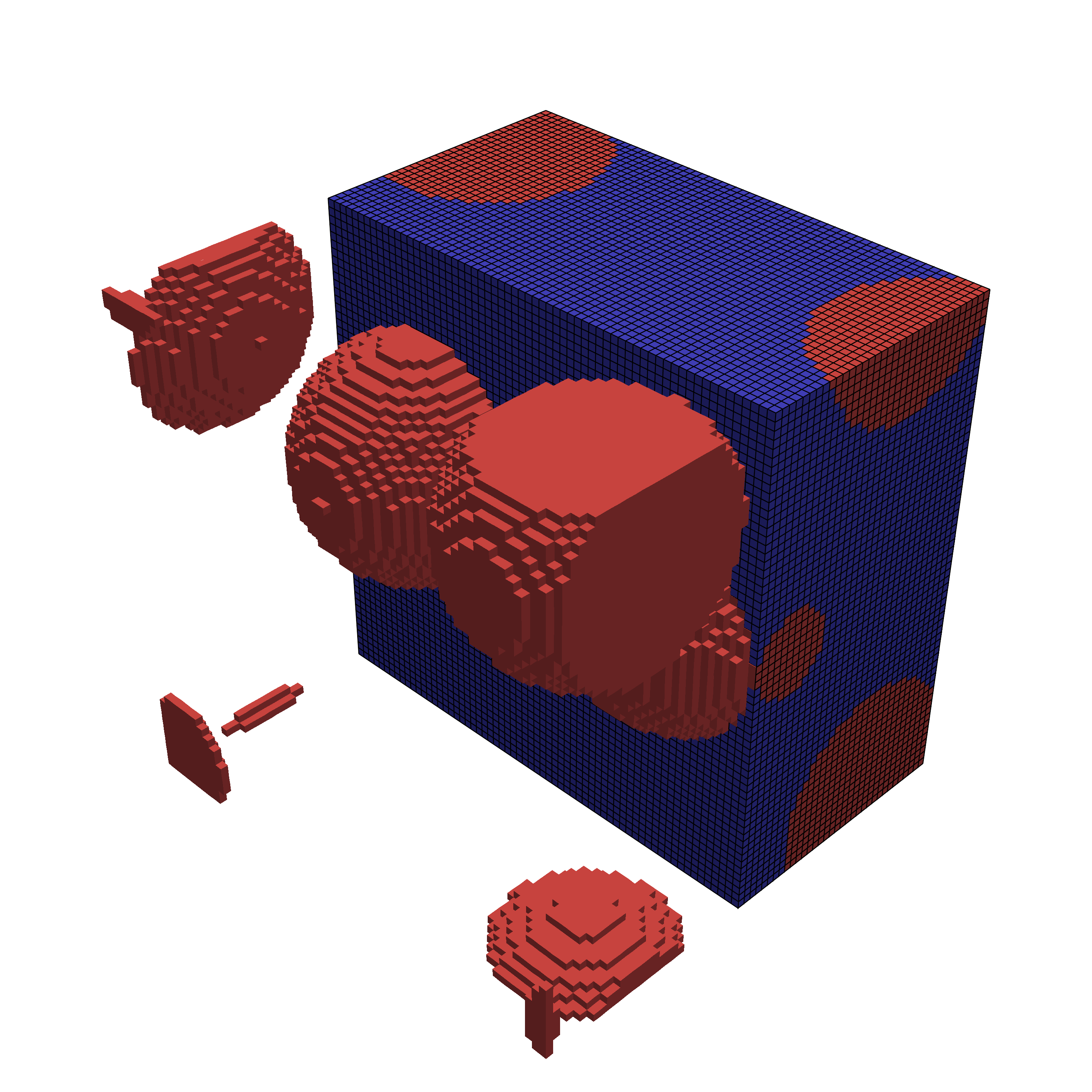}
\includegraphics[width=.32\textwidth]{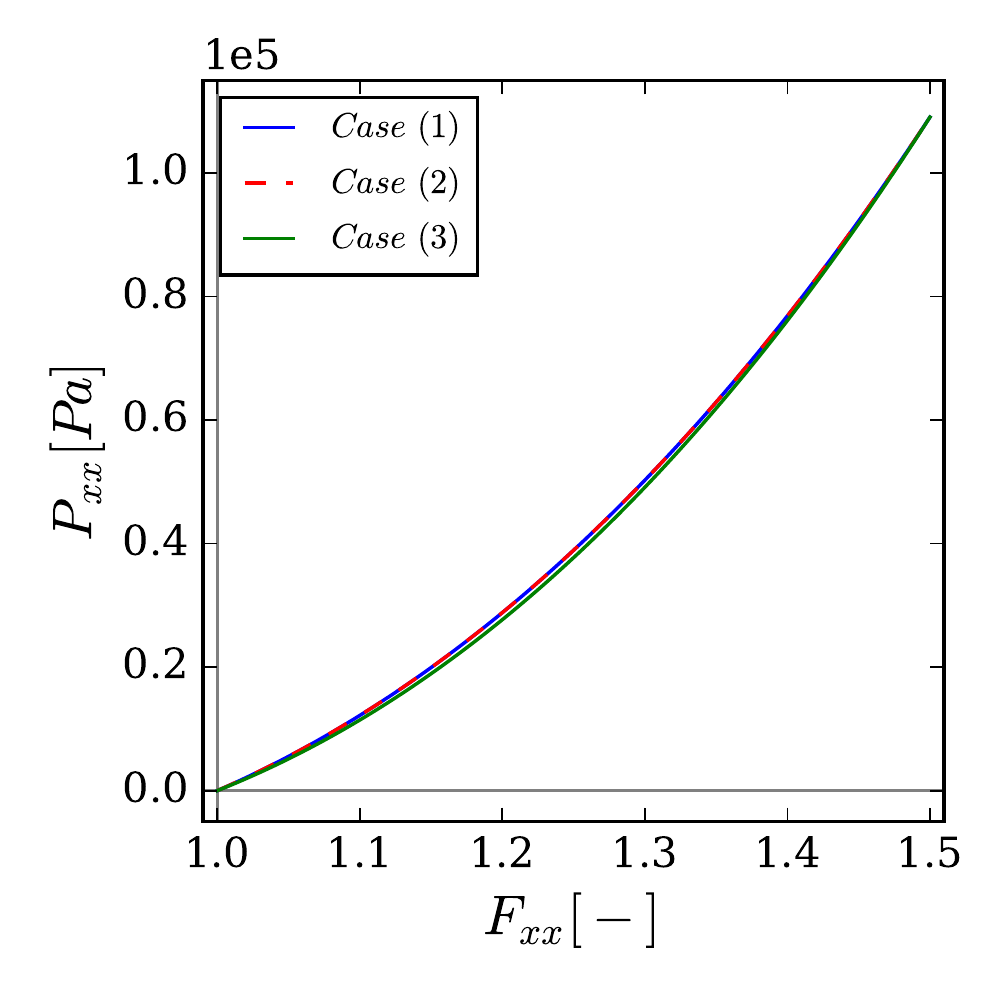}
\includegraphics[width=.32\textwidth]{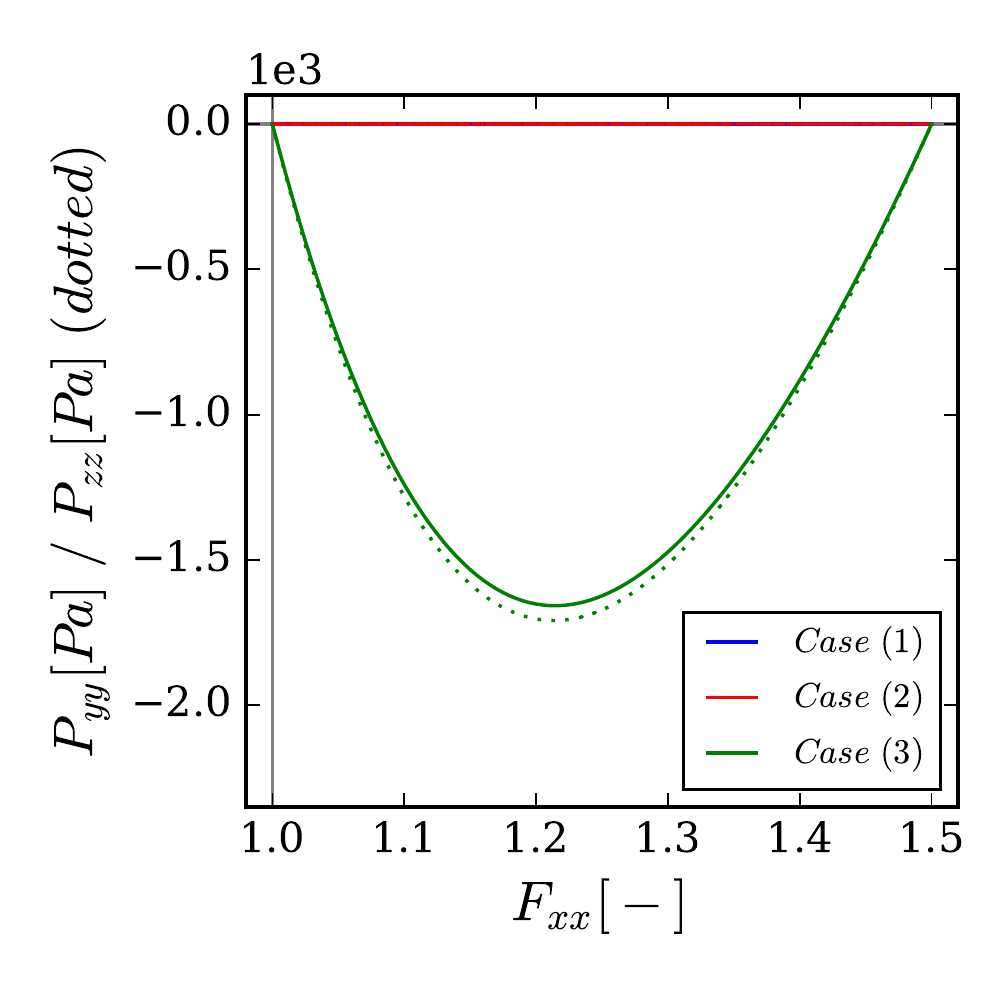}
\caption{Particle reinforced RVE (a), first Piola-Kirchhoff stresses for three load cases in hyperelastic model in load direction (b) and transverse directions (c).}
\label{fig1}
\end{figure}

The first benchmark consists in a non-linear elastic composite. A uniaxial monotonic test of a periodic cubic RVE composed by a hyperelastic matrix reinforced by stiff spherical particles will be simulated under macroscopic strain, stress and mixed control schemes. The RVE contains 4 particles occupying a volume fraction of 0.2 and the domain is discretized using $64^3$ voxels (Fig. \ref{fig1}(a)). The matrix and particle mechanical behavior follows a Saint Venant-Kirchhoff hyperelastic model used in \citep{Kabel2014,Geers2017}. The matrix properties are $\mu=28$Pa and $k=46.67$Pa (shear and bulk modulus) and the particles are one order of magnitude stiffer (both moduli are 10 times larger than the matrix ones). The uniaxial tensile test reaches an elongation of $\lambda_{end}=\frac{L}{L_0}=0.2$, being $L$ and $L_0$ the target and initial cell length in the loading direction, and being the other two directions unconstrained. This test is done using three different types of control that imply different loading paths. In all the cases the load is homogeneously divided in 100 load increments and equilibrium is solved in each of them. The load increments are defined introducing a pseudo-time $t$ being the final stage reached at $t=t_{end}$.

$Case\ (1)$ corresponds to the actual type of control in a uniaxial stress test, a mixed control in which a macroscopic deformation gradient is imposed in $x$ direction as a ramp and the rest of terms are stress free.
\begin{equation}
\begin{array}{lcr}
\overline{\mathbf{F}}^{(1)}\left(t\right)=\left(
\begin{matrix}
1+\lambda_{end}\frac{t}{t_{end}} & 0 & 0\\ 
*   & * & 0\\ 
*   & * & *
\end{matrix}
\right)
&
\ and \ 
&
\overline{\mathbf{P}}^{(1)}\left(t\right)=\left(
\begin{matrix}
*   & * & *\\ 
0   & 0 & *\\ 
0   & 0 & 0
\end{matrix}
\right)
\end{array}
\end{equation}
$Case\ (2)$ is a full stress control, being the stress applied a linear ramp from 0 to the stress $\mathbf{P}^{(1)}_{end}$ obtained as the solution in the first case and here introduced as input.
 \begin{equation}
\overline{\mathbf{P}}^{(2)}\left(t\right)=\mathbf{P}^{(1)}_{end}\frac{t}{t_{end}}
\end{equation}
Finally, $case\ (3)$ is a full strain control test (the standard boundary conditions in FFT) where all the deformation gradient is imposed using a linear ramp to reach the deformation gradient of previous simulation final solution $\overline{\mathbf{F}}_{end}^{(1)}$.
\begin{equation}
\overline{\mathbf{F}}^{(3)}\left( t \right)=\overline{\mathbf{F}}_{end}^{(1)} \frac{t}{t_{end}}
\end{equation}
The longitudinal response of the three test are represented in Fig. \ref{fig1}(b) while the transverse response is represented in Fig. \ref{fig1}(c).

All simulations have the same endpoint, but follow different paths. $Cases\ (1)$ and $(2)$ share the path, but differ in the size of the strain increments due to the non-linear behavior. Being the behavior of the hyperelastic material path independent, the macroscopic stress of $cases\  (1)$ and $(2)$ are superposed, as it can be observed in Fig. \ref{fig1}(b). $Case\  (3)$ follows a slightly different path because a ramp is applied controlling the full deformation gradient. In this material, the behavior depends non-linearly on the deformation gradient value, implying that the transverse elongation result of a transverse stress free condition depends on the deformation level. The effect of the different loading paths can be observed in the response in both directions being specially clear in the transverse direction (Fig. \ref{fig1} (b)) where under a stress and mixed control transverse stresses are zero while, when using strain control, small compressive stresses are developed. Macroscopic and microscopic results at the endpoint for all the cases are the same within the numerical accuracy.

\subsection{Polycrystal}
\begin{figure}[h]
\includegraphics[width=.32\textwidth]{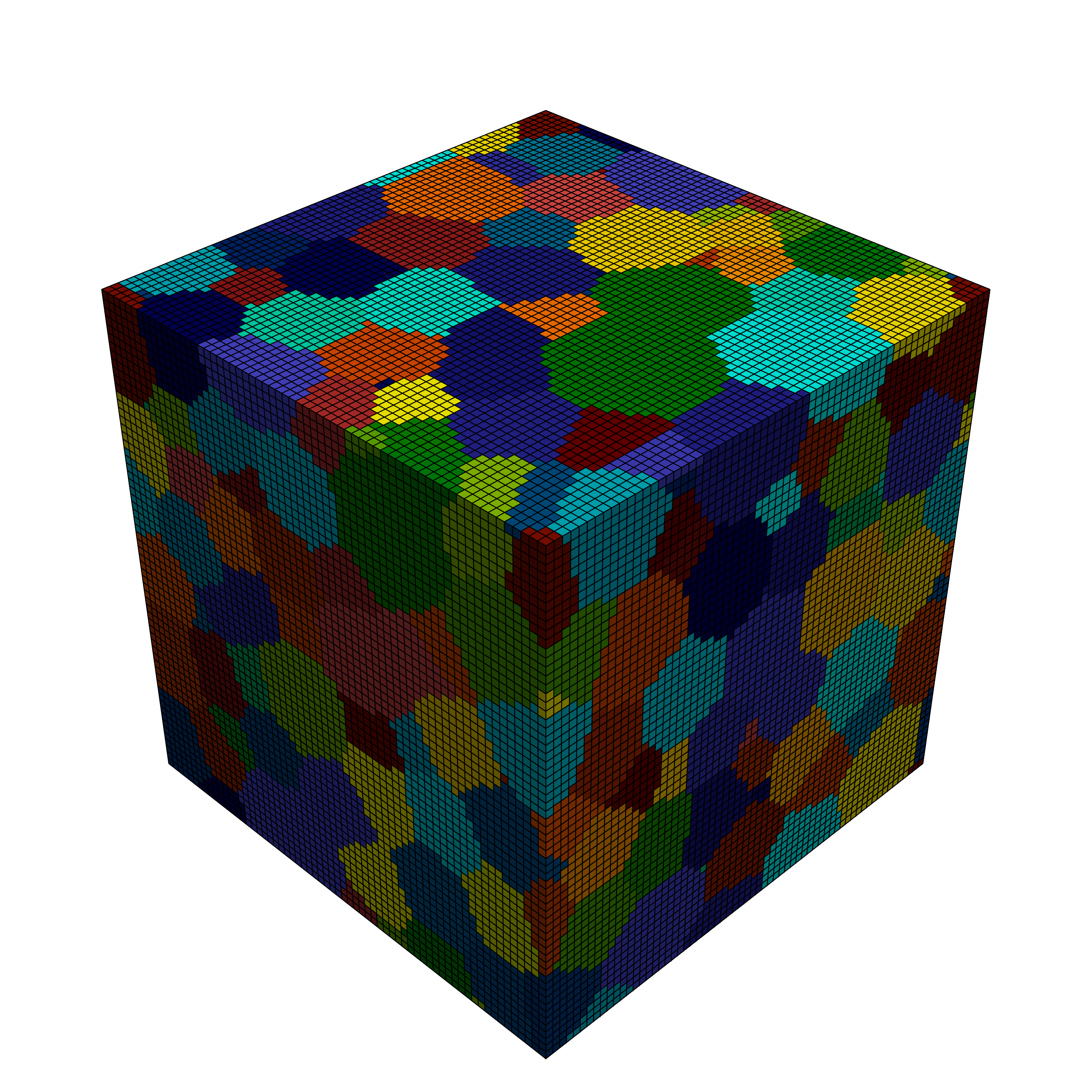}
\includegraphics[width=.32\textwidth]{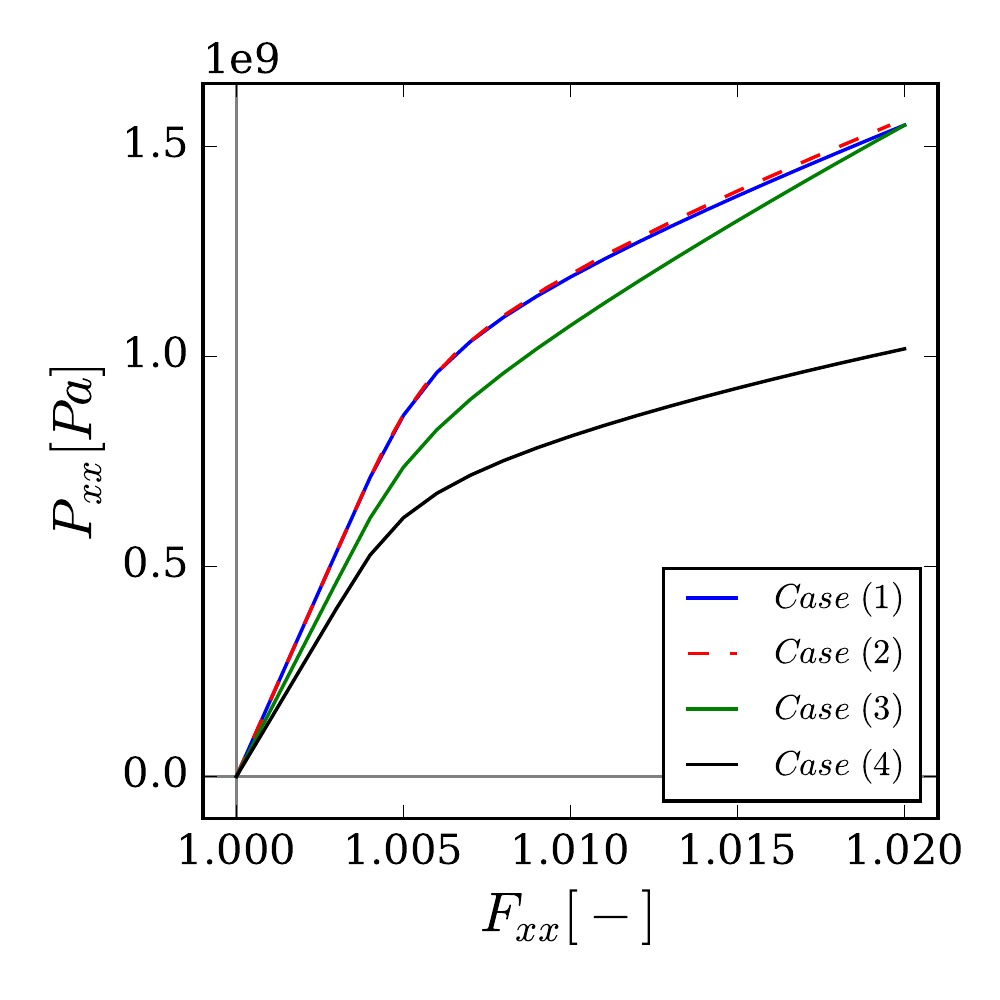}
\includegraphics[width=.32\textwidth]{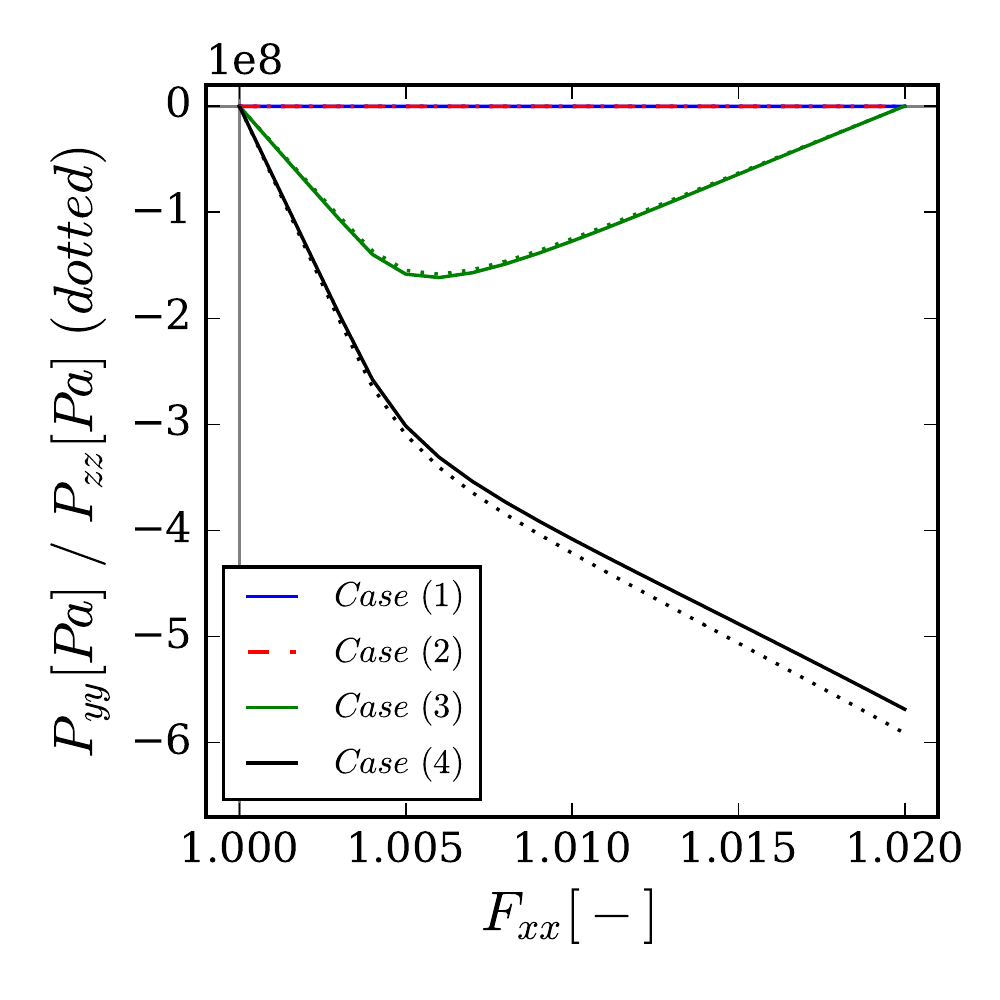}
\caption{Polycrystalline RVE (a), first Piola-Kirchhoff stresses for three load cases in crystal plasticity model in load direction (b) and (c) transverse directions.}
\label{fig:poly}
\end{figure}

The second benchmark is a polycrystal, an example with loading path and strain rate dependencies . The RVE is cube discretized in $64^3$ voxels and the model contains around 200 randomly oriented grains (Fig. \ref{fig:poly}(a)). Each grain follows a crystal plasticity elasto-visco-plastic behavior dependent on the orientation and on its internal variables. The details of the crystal plasticity model can be found in \citep{Lucarini2018}. A final elongation of $\lambda_{end}=0.02$ is imposed in 200s, what corresponds in the case of strain control to a uniaxial tensile test with strain rate $5\ 10^{-5}$ and time is discretized in 20 regular increments. Four loading conditions will be studied being the first three cases identical to the previous section. The fourth case consists in a \emph{pseudo-uniaxial} tensile test in which the uniaxial tension condition is approached using full strain control resulting in a isochoric deformation at the final stage. In the absence of an efficient stress-control technique this approach has been usually applied for homogenizing the response of nearly incompressible materials. The resulting deformation gradient follows
\begin{equation}
\overline{\mathbf{F}}^{(4)}\left(t\right)=\frac{t}{t_{final}}\left(
\begin{matrix}
\lambda_{end} & 0 & 0\\ 
0   & \left(\sqrt{\frac{1}{1+\lambda_{end}}}-1\right) & 0\\ 
0   & 0 & \left(\sqrt{\frac{1}{1+\lambda_{end}}}-1\right)
\end{matrix}
\right)+ \mathbf{I}
\end{equation}

The results of the simulations are represented in Fig. \ref{fig:poly}. Now $cases\ (1)$ and $(2)$, although very similar,  are not exactly superposed do to the effect of the differences in the strain rate. $Case\ (3)$ reaches the same point as previous two cases but, as it happened for the composite, the path is totally different and compressive transverse stresses are developed during the test. The $case\ (4)$ is a typical approximation made in plastic materials to perform tests emulating uniaxial stress, but controlling the full deformation gradient instead. However, it can be observed that for the properties here selected, the polycrystalline response (Fig. \ref{fig:poly}) is very different from a real uniaxial tensile test. This difference is specially noticeable when comparing the transverse response: in the \emph{pseudo-uniaxial} stress test,  the compressive transverse stresses are not zero and grow with the applied strain. This last result indicates that the use of this type of strain control conditions to emulate uniaxial tension can lead to inaccurate results. 

\section{Performance comparison}
\begin{figure}[h]
\includegraphics[width=.33\textwidth]{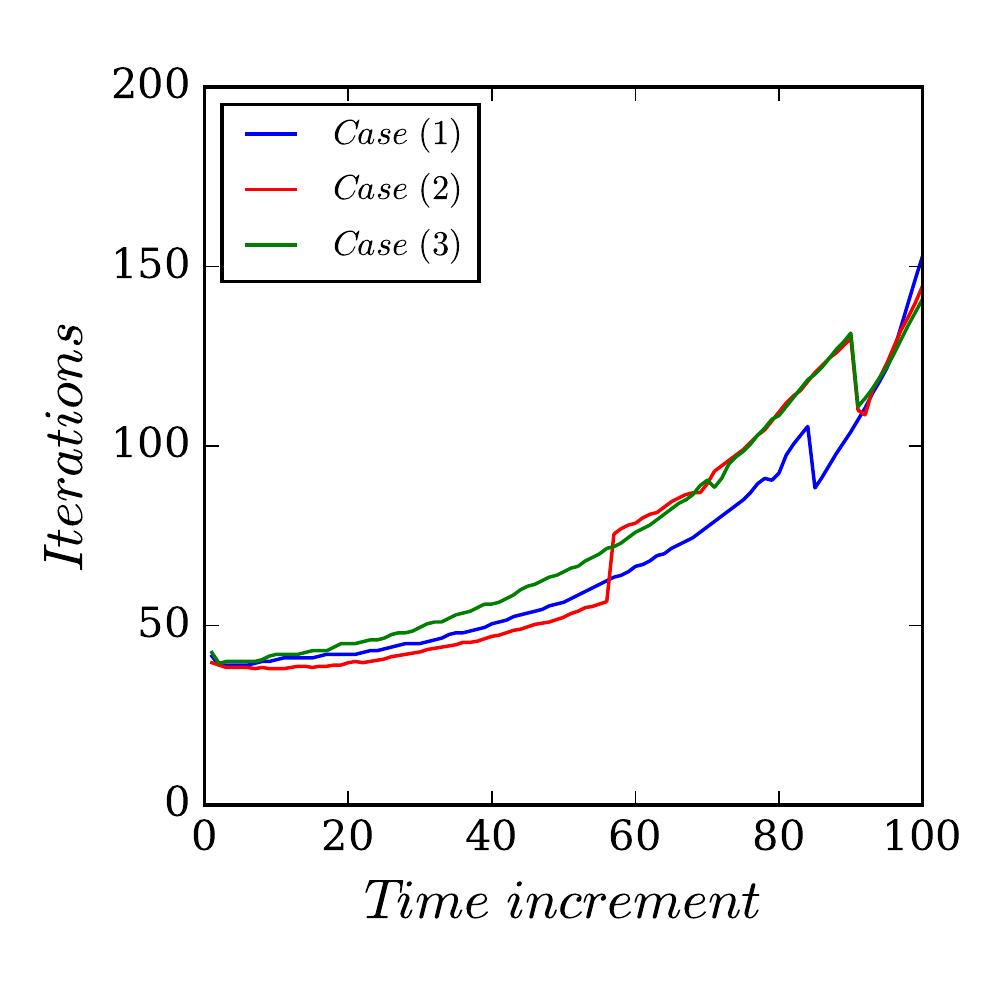}
\includegraphics[width=.33\textwidth]{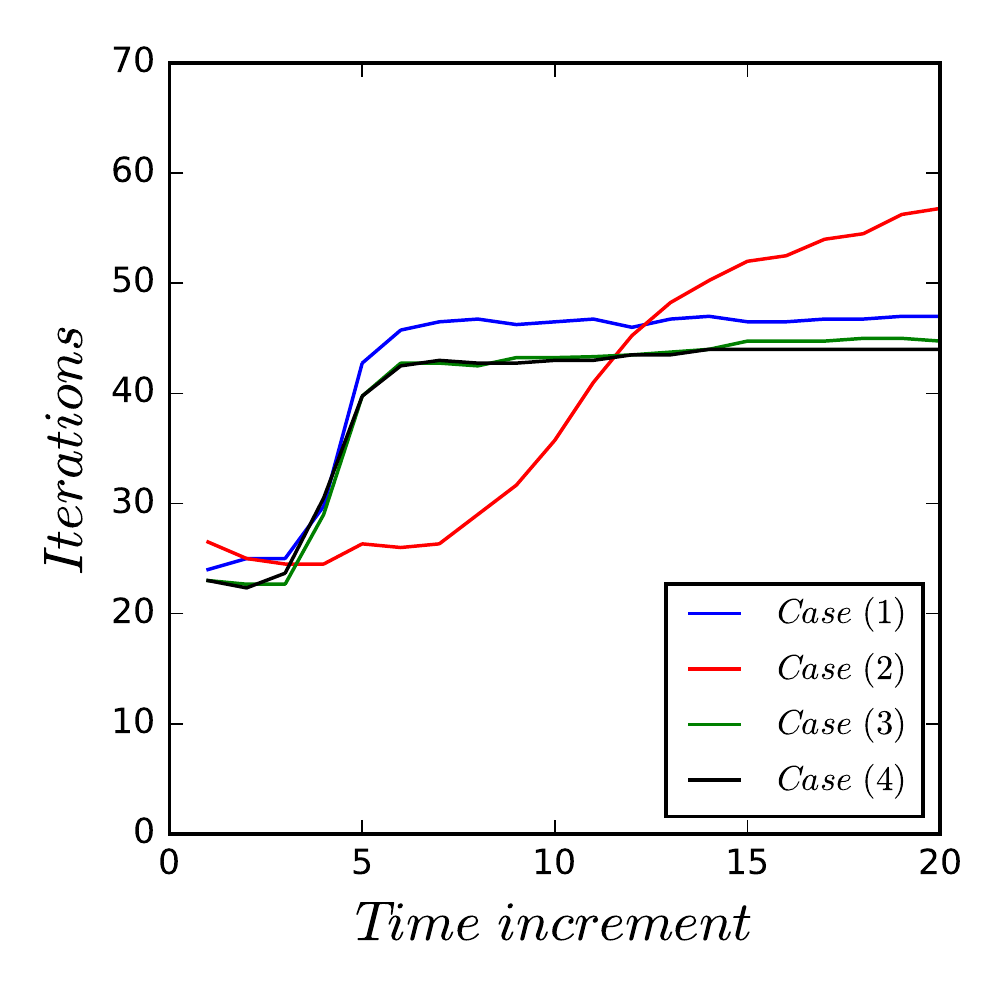}
\includegraphics[width=.33\textwidth]{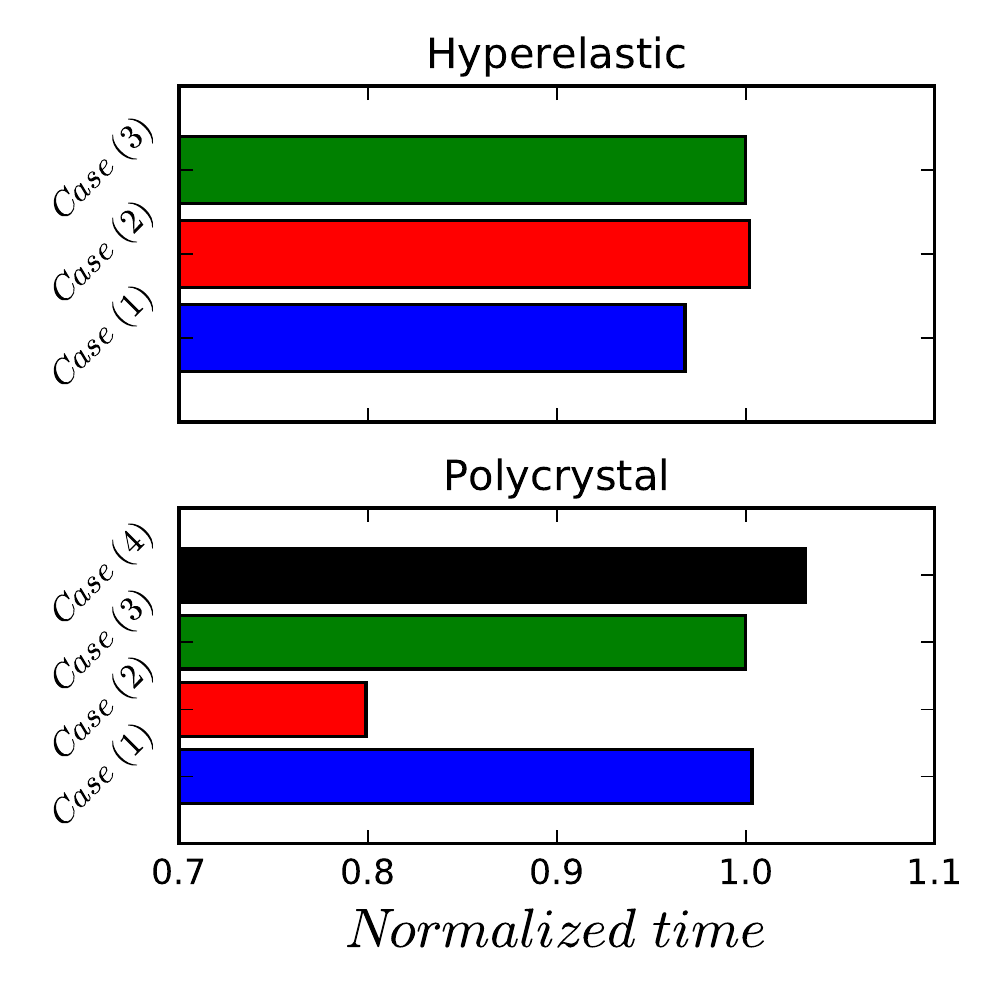}
\caption{Comparative of convergence for each time increment for (a) particle reinforced hyperelastic and (b) polycrystal. (c) Simulation time comparative for all examples normalized with respect to $case\ (3)$ time.}
\label{fig:conv}
\end{figure}

The performance of each type of test is analyzed here. The convergence criteria for the iterative linear solver (eq. \ref{linequil}) is the voxel averaged value of the $L_2$ norm of the residual and for the Newton solver is the maximum deformation gradient correction relative to the averaged deformation gradient. The tolerances used are  $10^{-8}$ and  $10^{-6}$ for linear and Newton solvers respectively. First, the total number of conjugate-gradient iterations per load increment is represented as function of the step (Fig. \ref{fig:conv}). This number includes the sum of all the linear iterations performed for each Newton iteration in an increment. In the case of the hyperelastic model (Fig. \ref{fig:conv}(a)) the stress and mixed controlled simulations ($cases$ 1 and 2) took approximately the same number of iterations than using full strain control, illustrating the very good performance of the approach proposed here. The jumps observed in the figure are due to the increase or decrease of one Newton iteration after a given time increment, because each Newton iteration implies several iterations for solving the linear system. In the case of the polycrystal (Fig. \ref{fig:conv}(b)) a similar tendency is achieved. The main difference between the different cases is that, when load control is applied using a ramp ($case \ (2)$), most of the load increments are elastic and solved in one Newton iteration corresponding the plastic regime only a few steps which require much higher number of iterations.
The comparative in simulation times for all the cases considered is represented in Fig. (\ref{fig:conv} (c)).  The hyperelastic simulations needed almost identical times for every boundary condition. On the contrary, in the polycrystal, differences are found for $case\ (2)$. As explained before, in this case most of the increments are purely elastic requiring very small simulation time. The plastic regime is then confined only in a few steps and, although each plastic step takes more time than in the other cases, the fast elastic steps compensate the global time and this load case is clearly faster than the others. It must be noted that this is a consequence of using a fixed time step and, for adaptive time stepping, all the cases will require very similar simulation times.
\section{Conclusions}
A new algorithm is proposed to impose a macroscopic stress or mixed stress/deformation gradient history in the context of non-linear Galerkin based FFT homogenization. This algorithm does not imply extra iterations compared to traditional schemes. The examples show the potential of the approach for two types of constitutive models, but can be used for any material. 

In summary, these algorithms represent a very useful tool for FFT based computational homogenization because it will allow to solve problems with mixed boundary conditions without extra computational effort and sustaining the well known advantages of the FFT methods for homogenization.

\section*{Acknowledgments}
This investigation was supported by \emph{ITP Aero} through the project MICROMECH II, the \emph{Comunidad de Madrid} through the grant PEJD-2016/IND-2824 and the \emph{Spanish Ministry of Economy and Competitiveness} through the project DPI2015-67667-C3-2-R.

\end{document}